\documentstyle[prl,aps,preprint]{revtex}

\begin{document}
\tighten

\title{HADRON SUBSTRUCTURE \\  PROBED WITH HADRON BEAMS\thanks
{This work is supported in part by funds provided by the U.S.
Department of Energy (D.O.E.) under cooperative agreement
\#DF-FC02-94ER40818.}%
\thanks{Talk presented at the NSAC/DNP town meeting on
``Nuclear Physics with Intermediate
and High-Energy Hadron Probes,'' Argonne, January 29--30, 1995.}}

\author{Xiangdong Ji }

\address{Center for Theoretical Physics \\
Laboratory for Nuclear Science \\
and Department of Physics \\
Massachusetts Institute of Technology \\
Cambridge, Massachusetts 02139 \\
{~}}

\date{MIT-CTP-2410 \hskip 1in  HEP-PH/9502265
 \hskip 1in January 1994}

\maketitle

\begin{abstract}
In this talk, I focus on the quark-gluon structure of hadrons probed
using high-energy hadron beams.  I start with a brief review on recent
major achievements in measuring parton distributions of the nucleon,
pion, and kaon, with hadron facilities at CERN and FNAL\@. Then I
discuss a number of outstanding questions and interesting physics
issues in the field, and point out their intellectual impact on
nuclear physics as a whole.  While advocating a continuing
exploitation of hadron beams at CERN and FNAL, I strongly emphasize
the role of a polarized RHIC, where a major nuclear physics program on
the structure of hadrons can thrive.
\end{abstract}

\pacs{xxxxxx}

\section{introduction}

Study of the quark-gluon structure of hadrons, the nucleon in
particular, has becoming a major frontier in modern nuclear
physics. Since mid-1970's, a large number of theoretical nuclear
physicists have involved and played a major role in understanding the
substructure of hadrons. In experimental nuclear physics community, a
large amount of resources has been devoted to measuring properties of
the nucleon that have simple and direct attribution to its
substructure.  One of the most visible activities in recent years has
been measuring and interpreting the spin structure functions of the
nucleon in deep-inelastic scattering~\cite{G1}.

It is simple to understand why the study of hadron
structure is an exciting and promising field of nuclear physics.
Since Hofstadter's measurement of the nucleon's size 1950's, its inner
structure and dynamics have become a major focus
for nuclear and particle physicists. Not only the nucleon is the
basic constituent of the nucleus, and thus knowing its properties is
crucial for solving the nuclear systems, but also
it is the king of hadrons, and unlocking its secrets is
an important milestone in exploring the hadron world.
In 1960's, it was thought that
the structure of the nucleon is just another incarnation of
shell model, which had successfully explained the structure of
atoms and nuclei.  However, the MIT-SLAC deep-inelastic scattering experiments
and the advent of Quantum Chromodynamics (QCD) had forced us
to abandon the simplistic point of view and to face
the full complexity of a relativistic quantum field theory.
In the past twenty years, we have learned a lot about QCD
and hadron physics, however, the problem of hadron structure
is not completely solved. Before theorists
can calculate as confidently as they did for QED, experimental
probes into the structure of hadrons continue to play a
pivotal role in our quest.

The goal of this talk is to review accomplishments made in this field
using high-energy hadron beams and to point our their future prospects
in solving outstanding questions. Before I plunge into details, I
would like to point out some important virtues of hadron beams from a
theorist point of view. First of all, hadron beams are effectively
beams of quarks and gluons.  There are variety of hadron beams:
proton, antiproton, pion, kaon, etc., which can be used for different
studies. Second, there are many hard scatterings which one can select
for probing different aspects of hadron structure, e.g., jet, direct
photon, Drell-Yan, heavy mesons, and weak bosons ($W $and $Z$)
production.  Lastly, experimenters can now produce high-energy proton
beams with polarization, which allows us to study important
spin-dependent parton distributions.

\section{MAJOR ACCOMPLISHMENTS OF THE PAST}

Direct probe of the quark-gluon content of hadrons with hadron beams
started with low-energy facilities, such as $J/\psi$ production at
AGS, in the 70's. However, experiments at these facilities are barely
explainable with perturbative QCD, and the statistical accuracy of the
data does not allow a meaningful extraction of parton distributions.
Targeted studies of quark and gluon distributions at hadron facilities
began in the 80's at CERN SPS and FNAL Tevatron.  With beam energies
of 200 GeV and higher, there is little question that hard subprocesses
are perturbative and radiative effects are calculable in perturbation
theory.  For fixed target experiments, important observables include
Drell-Yan pairs, direct photon, and heavy-quarkonium production. In
the simple parton model, they are directly related to anti-quark and
gluon distributions, which are less constrained by deep-inelastic
scattering data.  In the following subsections, I briefly review the
main results of these studies.  I apologize for not being complete.

\subsection{Direct-photon production}

Direct photon production provides a sensitive probe to the gluon
distribution in hadrons.  From deep-inelastic scattering (DIS) data,
it was determined indirectly that gluons carry 50\% of the nucleon's
momentum in infinite momentum frame~\cite{DIS}.  For certain very high
energy experiments, the gluon content of a hadron plays a more
dominant role than the quark content. Thus an accurate measurement of
the gluon distribution is extremely important.  In DIS process, the
total cross section depends on gluons only at the next-to-leading
order. However, direct photon production through Compton process $g
q\to \gamma q$ involves the gluon distribution starting at the tree
level.

Many experiments have measured direct photons in proton-induced
collision~\cite{DP}.  Here I would like to mention particularly the
WA70 experiment at CERN SPS, in which prompt photons with $p_T$ in the
range of 4.0 to 6.5 GeV and $x_F < 0.45$ were measured.  The result is
shown in Fig.~1 for different $x_F$ ranges.  The solid and dashed
curves correspond to two versions of the gluon distribution by Duke
and Owens~\cite{DO}, which were obtained by fitting data from
deep-inelastic scattering, the Drell-Yan process, and $J/\psi$
production. Clearly the data is discriminatory and favors the ``soft''
gluon distribution,
\begin{equation}
          xG(x) = (1+9x) (1-x)^6 \ ,
\end{equation}
at $Q^2=4 {\rm GeV}^2$ with $\Lambda_{\rm QCD}
= 200$ MeV\@. Of course, one shall not be
satisfied with testing certain parameterizations.
The data shall be used for global fits~\cite{TUNG}.

$J/\psi$ production can also constrain the gluon
distribution. However, it has limitations. For instance, a model for
the soft process $c\bar c\to J/\psi$ must be used to connect the cross
section with the gluon distribution.

\subsection{Drell-Yan process and the sea quark distributions
in the nucleon}

In Drell-Yan process, quarks and antiquarks from hadron beams and
nuclear targets annihilate each other and produce virtual photons that
subsequently decay into lepton pairs ($l^+l^-$).  A great virtue of
the process is that it depends on valence and sea distributions
separately. Through a combination of experiments with selected
kinematics, one can measure sea quark distributions accurately. Of
course, combining parity-conserving and parity-violating structure
functions from neutrino DIS, one can also extract the quark
sea. However, Drell-Yan data provides not only a cross check, but also
a better alternative in certain cases.

There are a number of interesting questions that one can ask about the
sea quark distributions in the nucleon: Are the sea distributions
measured in Drell-Yan scattering consistent with that determined from
neutrino DIS? Is the up-quark sea in the nucleon the same as the
down-quark sea in the nucleon? Are the sea-quark distributions
modified in a nuclear environment?  These questions have been
partially answered with a number of excellent experiments.

Three experiments have made comprehensive measurements of large mass
$\mu^+\mu^-$ pairs in proton-induced collisions: E288 \cite{E288} and
E605 \cite{E605} experiments at FNAL, and NA3 \cite{NA3} experiment at
CERN\@. An example of the data is shown in Fig.~2. With the
order-$\alpha_s$ perturbative QCD corrections, the data is consistent
with the anti-quark distribution extracted from neutrino
scattering. Together, they provide accurate constraints on the sea
quark distributions. One interesting point here is that both E288 and
E605 data show a positive slope for the cross section at $y=0$ for
(see Fig.~3). Since the Cu target contains more neutrons than protons,
it is natural to conclude that $\bar u$ is suppressed relative to
$\bar d$ in the proton.

Actually, the assertion that the up-quark sea
is not the same as the down-quark sea is also a natural
explanation for the violation of the
Gottfried sum rule~\cite{NMC},
\begin{eqnarray}
     \int^1_0{(F_2^p(x) - F_2^n(x)) \over x}dx
     &=&  {1\over 3}+{2\over 3}\int^1_0dx[\bar u(x) -\bar d(x)]  \nonumber \\
     &=&  0.240\pm 0.016 \ ,
\end{eqnarray}
which is measured by EMC at $Q^2 = 4 \hbox{ GeV}^2$.
To get a decisive evidence that
up and down seas are not symmetric, Garvey {\it et al}.~proposed
to compare the Drell-Yan cross section in $p+p$
and $p+d$ collisions~\cite{GARVEY}. A recent experiment
at CERN \cite{NA51} has measured one
data point at $x=0.18$, where $\bar u/\bar d=0.51\pm 0.04\pm 0.05$.
This is a first direct evidence that the quark sea is
not isospin singlet.

Motivated by the EMC effect that quark distributions in a nucleus are
different from these in a nucleon~\cite{EMC}, the E772 experiment
\cite{E772} at FNAL measured for the first time the nuclear dependence
of sea quark distributions through Drell-Yan $\mu$-pair production.
In the $x$ range $0.1\sim 0.3$, they found that the sea quark
distributions in measured nuclei are the same as that in a nucleon
(see Fig.~4). Below $x=0.1$, the sea-quark shadowing is
discovered. The effects, however, are less pronounced than the
combined valence and sea shadowing measured by EMC (modulo $Q^2$
evolution).  The result implies that the momentum fraction carried by
sea quarks in a nucleus is reduced through shadowing. The result also
rules out certain models that are motivated to explain the EMC
effects.  This high-precision, definitive experiment is very
impressive.

\subsection{Parton distributions in the pion and kaon}

The parton distributions in the pion cannot be measured in
deep-inelastic scattering. The only probe is through high-energy
$\pi-N$ scattering, detecting Drell-Yan pairs and direct photon
production. The valence distribution in the pion was first
investigated by E444 collaboration at Fermi Lab. High statistics and
larger kinematic coverage experiments were later done by NA3 and NA10
collaborations at CERN, E615 collaboration at FNAL.

The valence quark distribution is well-determined from the data
measured in NA10 \cite{NA10} and E615 \cite{E615} experiments. The
$x\to 0$ and $x \to 1$ behavior of the distribution is consistent with
expectations of Reggie theory and the quark counting rule.  The gluon
distribution can be extracted from the direct photon production data
from the WA70 experiment~\cite{PIW70}. The sea quark distribution is
less well-determined, primarily because the data only exists in the
$x_\pi> 0.2$ region. A recent fit to these experiments by Sutton,
Martin, Roberts, and Stirling are shown in Fig.~5~\cite{SMRS}. One
interesting consequence of the data is that the color fields in the
pion approach that in the QCD vacuum in the chiral limit, a feature
that is consistent with the postulate that the pion is a collective
excitation of the QCD vacuum~\cite{XDJ}.

Higher-twist effects were observed in the angular distribution of the
Drell-Yan pairs. At the leading twist level, the angular distribution
goes like $1+\cos^2\theta$. However, as $x_\pi \to 1$, that is, when
the valence quark carries all the momentum of the impinging pion, the
experimentally-observed distribution goes like
$1-\cos^2\theta$~\cite{HEIN} (see Fig.~6). This is a clear indication
that the twist-four effect (multi-parton coherent scattering) plays a
dominant role~\cite{BRAN}.

The parton distribution in the kaon was measured by NA3
collaboration. The data indicates that the up quark distribution in
the kaon vanishes quicker than that in the pion. More data is needed
to get a more complete picture of parton distributions there.

\section{Outstanding questions}

As I said in the beginning, our theoretical knowledge about hadron
structure is very limited.  No one knows how to calculate parton
distributions from the first principle.  Various hadron models have
been invented, which have their own limitations.  One of the
limitations is that models use effective degrees of freedom, not those
of QCD\@. Thus it is very difficult to do a meaningful computation of,
e.g., gluon distribution in a model. Ultimately, our goal is not to
test models --- we want to test QCD\@.

Before one can come up with a realistic picture of the nucleon, it is
important to collect as much data as possible. The advent of BCS
theory for superconductor would not be possible if without some
crucial experimental data. In the case of hadron structure, one
important piece of information is the spin and flavor composition of
parton distributions.  From QCD, one can make systematic
classifications of parton observables in a hadron~\cite{JI1}. In the
following subsections, I discuss some outstanding questions from the
point of view of this classification.

\subsection{Unpolarized Distributions}

With three quark flavors, there are seven unpolarized,
leading-twist distributions in a hadron,
\begin{equation}
        u(x), \ \bar u(x), \ d(x), \ \bar d(x), \ s(x) , \ \bar s(x),
        G(x) \ .
\end{equation}
Some of these have been well measured in DIS and hadron-beam induced
reactions. Some distributions, however, are less known.  Some open
questions are suitable for study with hadron beams.

\subsubsection{$u$ and $d$ sea distributions in the nucleon}
We know fairly well the average of up and down sea distributions.
However, the separate distributions are not known.  Sea distributions
are an important window to the hadron structure. Many people believe
that most of the sea quarks arise from the $Q^2$ evolution in which
quark pairs are created through splitting of gluons. On the other
hand, sea quarks at low energy scales (intrinsic sea) are certainly
related to low-energy and long-distance dynamics.  The difference
between up and down sea distributions will put many ideas, including
Pauli-blocking effects and pion cloud, into test.

\subsubsection{Strange-antistrange distributions in the nucleon}
The most recent measurement of the strange quark distributions was
done by the CCFR collaboration~\cite{CCFR}. Within the precision of
the data, it seems that $s(x)$ and $\bar s(x)$ distributions are the
same. However, there are reasons to believe $s(x)-\bar s(x)$ is
nonzero. If so, the size of this difference is very interesting and
shall be measured. According to the picture of meson cloud, the
valence anti-strange quark in virtual kaons present in the nucleon
wave-function contributes to $\bar s$ distribution, and the valence
strange quark in virtual hyperons contributes to $s$
distribution. With this mechanism, there is no symmetry between
strange-anti-strange distributions. Since the total flavor charge is
conserved,
\begin{equation}
      \int^1_0 (s(x) -\bar s(x)) = 0\ ,
\end{equation}
the size of $s(x)-\bar s(x)$ tells us the locality of
quark-antiquark pairs in momentum space, a
important observable of the sea.

\subsubsection{Parton distributions in the kaon}
One can learn a lot by comparing the parton distributions in the pion
and kaon. Both particles are Goldstone bosons, however, the kaon is
much heavier than the pion. The effect of the strange quark mass in
the kaon structure is valuable in studying structure of Goldstone
bosons.

\subsection{Polarized Distributions}

With recent measurements of the longitudinally-polarized quark
distributions at CERN and SLAC~\cite{G1}, high-energy spin physics has
becoming a hot subject.  Theoretical studies revealed that there are a
number of interesting spin-dependent quark-gluons distributions which
have never been measured before and which have extremely important
implications about the spin content of the
nucleon~\cite{JAFFEJI}. Most of these distributions, however, cannot
be measured cleanly in lepton-nucleon deep-inelastic scattering either
due to chirality selection rule or due to small contribution in
lepton-induced processes. On the other hand, a polarized hadron beam
is far more superior for studying these observables.  Let me highlight
some important physics one can hope to study with a polarized proton
beam.

\subsubsection{Polarized gluon distribution $\Delta G(x)$}
In a polarized nucleon, gluons are also polarized. The polarization
contributes to the nucleon spin=1/2. Normally, one would expect the
contribution to be some fraction of the total. However, some advocate,
according to a study of the axial anomaly~\cite{AR}, that the gluon
polarization contributes several units of angular momentum. If so,
there must be mechanisms to cancel this large gluon-spin
contribution. However, before one starts to look for the cancelation,
one must measure $\Delta G(x)$ in the full $x$ range and study its sum
rule. So far, however, nothing is known about $\Delta G(x)$
experimentally.

\subsubsection{Polarized quark sea}
Another interpretation of the ``spin crisis'' is that the sea quarks
have large polarization. This is deduced from the hyperon
$\beta$-decay data plus use of the flavor SU(3) symmetry. One can
verify this deduction through measuring the polarized distributions
for up, down, and strange seas independently. Such data, if available,
will test SU(3) symmetry, very important from the point of hadron
structure.

\subsubsection{Transversity distribution $h_1(x)$}
For a spin-1/2 nucleon, there are a total of three distributions which
characterize the quark state in leading order high-energy scattering:
the unpolarized quark distribution and quark helicity distribution we
have talked about above, and the quark transversity distribution. The
last distribution shows up in a transversely polarized nucleon and
counts the net number of quarks polarized along the transverse
polarization of the nucleon.  In non-relativistic quark model or in
models with no parton interactions, the transversity distribution is
the same as the helicity distribution.  Jaffe and I have derived a sum
rule for the quark distributions~\cite{JAFFEJI},
\begin{equation}
      \int^1_0 h_1^q(x, Q^2) dx = \delta q
\end{equation}
where $\delta q$ is the tensor charge of the nucleon. A recent
estimate indicates that $\delta u = 1.0 \pm 0.5$ and $\delta d = 0.0
\pm 0.5$~\cite{HJ}. A measurement of the tensor charge is important to
understand the relativistic effects and the spin structure of the
nucleon.

\subsubsection{Twist-three parton distributions}
There are a number of twist-three distributions which are
interesting. The most famous one is $g_2(x)$, present in a
longitudinally polarized nucleon and measurable in deep-inelastic
scattering~\cite{G2}. Another distribution similar to $g_2(x)$ is
$h_2(x)$, present in a transversely polarized
nucleon~\cite{JAFFEJI}. There are also general twist-three
distributions which depend on two Feynman variables~\cite{Sterman}.
Higher-twist distributions appear in coherent parton scattering, which
is not present in Feynman's parton model.  All twist-three
distributions can be accessed in hadron-beam induced processes.

\section{Future experimental prospects}

To probe directly the quark-gluon constituents of the nucleon, one
must ensure the probes are clean. That means one must resort to hard
processes in which quark and gluon scattering is perturbative and can
be calculated in perturbative QCD\@. As a consequence, we need hadron
beams at several hundred GeV\@.  At present time, such facilities are
limited to SPS at CERN, Tevatron at FNAL, and possibly HERA at
DESY\@. In the near future, RHIC will be added to the list. In the
last few years, we have witnessed increasing involvement at these
facilities by nuclear experimentalists. Such involvement shall
continue with clear physics goals and high-quality experiments. In the
following, I will discuss some prospects for the two US facilities.

\subsection{Tevatron at FNAL}

As I have discussed, the hadron beams at FNAL have already played
important roles in probing parton distributions. I think the nuclear
community shall continue to use these beams to make further
explorations of some of the outstanding questions. Some studies can
only be made with hadron beams, for instance, the quark and gluon
distributions in pions and kaons. Kaon beams have the unique
capability to probes the strange sea in the nucleon. Unfortunately,
because they are secondary beams, the intensity may not be suitable
for high-precision measurements.

One excellent example for future experiments at FNAL is E866
experiment proposed by G. Garvey {\it et. al.}, which was approved in
December 1992 and will start to take data in the beginning of the next
year. According to the proposal~\cite{E866}, the experiment makes a
precise measurement of Drell-Yan yields from hydrogen and deuterium.
The ratio of these yields is used to infer the ratio $\bar u(x)/\bar
d(x)$ in the proton, over the $x$ interval between 0.03 to
0.3. According to the parton model, one has,
\begin{equation}
            2{\sigma_{\rm DY}(p+p) \over \sigma_{\rm DY}(p+d)}
        = 1 - {\bar d(x) - \bar u(x) \over \bar d(x) + \bar u(x)} .
\end{equation}
The expected statistical accuracy together with several
model predictions for the ratio is shown in Fig.~7~\cite{PENG}.
Clearly, the experiment will make a first precise measurement of the
asymmetric sea and the data provides a definitive test of
various models.

\subsection{RHIC}

Although RHIC is not build for learning hadron structures, it turns
out that it offers excellent opportunities for such studies. The most
exciting possibility is of course a polarized RHIC\@. Thanks to an
ingenious invention called Siberian snake~\cite{KRISH}, one can now
have a polarized hadron beam at few hundred GeV\@. With 250 GeV
longitudinally or transversely-polarized proton beams colliding at
luminosities up to $2\times 10^{32}$ cm$^{-2}$sec$^{-1}$, one can make
state-of-art studies of polarized distributions in the proton.

Even without polarization, one can already do some interesting physics
with RHIC hadron beams. For instance, two recent studies
\cite{SOFFER,PEN2} show that by comparing W and Z production from
$p+p$ and $p+d$ collisions, one can extract the up and down quark
distributions separately. These measurements can provide a cross-check
on results from E866 measurements and extend them to smaller $x$
region.

The RHIC spin collaboration was formed three years ago to study the
feasibility of a polarized RHIC and to make proposals for physics
program at such machine~\cite{RSC}.  According to the proposal, the
major components for the acceleration of polarized beams to RHIC top
energy are shown in Fig.~8. Polarized protons are produced at the
present AGS source and are accelerated to 200 MeV with the
LINAC\@. They are captured in the AGS Booster and are further
accelerated to 1.5 GeV and then transferred to AGS where they are
accelerated to 25 GeV\@. A partial Siberian snake is needed to
maintain the polarization at AGS\@. When protons are transferred to
RHIC for acceleration to 250 GeV, two Snakes are needed for correcting
the depolarization effects. In a steady running, one expects 70\% beam
polarization.

The two RHIC detectors, PHENIX and STAR, will be used to measure
Drell-Yan pairs, direct photons, jets, $W$ and $Z$, and heavy meson
production. With these probes, one can make systematic measurements of
polarized quark and gluon distributions in the nucleon.

1).~~{\it Gluon helicity distribution\/}
can be measured in jet
and direct photon production. For low transverse momentum jets,
e.g.~$10<p_T<20$ GeV/c, single and di-jet productions
are dominated by gluon-gluon fusion. The double spin asymmetry
for longitudinally-polarized collision is
\begin{equation}
         A_{LL} = {\Delta G(x_1) \over G(x_1)}
       {\Delta G(x_2) \over G(x_2)} a_{LL}(gg\to gg) \ ,
\end{equation}
where $a_{LL}$ is the asymmetry at the parton level. The range of $x$
covered at RHIC extends at least from 0.05 to 0.3.  Direct photons are
produced through $q\bar q$ annihilation and $q-g$ Compton process. The
latter is dominant in polarized scattering and produces the following
asymmetry,
\begin{equation}
        A_{LL} = {\Delta G(x_1) \over G(x_1)}
       {\Delta q(x_1) \over q(x_1)} a_{LL}(qg\to q\gamma ) \ ,
\end{equation}
where the valence quark distributions can be taken from
DIS data.

2).~~{\it Polarized sea quark distributions\/}
can be measured through
Drell-Yan and weak bosons ($W$ and $Z$) production~\cite{DRSY}. In
Drell-Yan process, the double spin asymmetry is,
\begin{equation}
        A_{LL} =  a_{LL}(q\bar q\to l^+l^-)
      {\sum_f e_f^2 [\Delta \bar q(x_1)\Delta q(x_2)
   + \Delta \bar q(x_2)\Delta q(x_1) ] \over \sum_f e_f^2 [\bar q(x_1)q(x_2)
   + \bar q(x_2) q(x_1) ]                         }
\end{equation}
Without the sea quark polarization, the asymmetry vanishes.  For weak
boson production, one can define both single-spin and double-spin
asymmetry. The single-spin asymmetry violates the parity.  Both
symmetries are sensitive to sea-quark polarizations.

3).~~{\it The quark transversity distribution\/}
can be measured similarly
in Drell-Yan and Z-boson production~\cite{SOFFER,JI2}. Since the interference
between left and right-hand annihilations is required, $W$
production gives a vanishing asymmetry.

4).~~{\it Twist-three quark-gluon and gluon-gluon correlations\/}
can be measured through single-spin asymmetry associated
with pion production, single-jet production etc.

To sum up, hadron beams have helped us to learn a great deal about the
quark and gluon distributions of hadrons. They will continue to play
an important role in our future research in the field. A polarized
RHIC collider offers a unique and interesting physics opportunity to
the nuclear community.

\acknowledgments
I thank G. Bunce, G. Garvey, and J. C. Peng for informative discussions.

\begin{figure}
\bigskip
\caption{Invariant cross section for $pp\to \gamma X$. a)
$-0.35<x_F<-0.15$. b).$-0.15<x_F<0.15.$ c). $0.15<x_F<0.45$.
d). $-0.35<x_F<0.45$ from WA70 experiment. The dashed
and solid curves represent
the prediction from Duke and Owns hard and soft gluon distributions. }
\label{fig1}

\bigskip
\caption{Drell-Yan cross section measured by NA3 and E605 collaborations.
\qquad\qquad\qquad~}
\label{fig2}

\bigskip
\caption{Drell-Yan cross section measured by NA3 and E605 collaborations,
plotted as a function of the rapidity of virtual photons.}
\label{fig3}

\bigskip
\caption{Ratios of the Drell-Yan dimuon yield per nucleon for positive
$x_F$ from E772 collaboration.}
\label{fig4}

\bigskip
\caption{Valence, sea, and gluon distributions obtained by
Sutton {\it et al}.~through fitting NA10, E615, and W70 experiments.}
\label{fig5}

\bigskip
\caption{The parameter $\lambda$, as defined in the angular
distribution of the Drell-Yan pairs, $1+\lambda\cos^2\theta$,
extracted from E605 experiment.}
\label{fig6}

\bigskip
\caption{Predictions of Drell-Yan cross section ratios for
various models. The expected sensitivities for E866
and the recent N51 results are also shown (from Ref.~\protect\cite{PENG}).}
\label{fig7}

\bigskip
\caption{Polarized proton collision at Brookhaven.
\qquad\qquad\qquad\qquad\qquad\qquad\qquad\qquad\quad\enspace}
\label{fig8}

\end{figure}

\end{document}